\documentclass[structabstract]{aa}  
\usepackage{graphicx}
\usepackage{txfonts}
\usepackage{amssymb}
\begin{document}
   \title{Obscuration of Supersoft X-ray Sources by Circumbinary Material}

   \subtitle{A Way to Hide Type Ia Supernova Progenitors?}

   \author{M.T.B. Nielsen
          \inst{1}
          \and
          C. Dominik\inst{1,2}%\fnmsep %\thanks{Just to show the usage of the elements in the author field}
	  \and
	  G. Nelemans\inst{1,3}
	  \and
	  R. Voss\inst{1}
          }

   \institute{Department of Astrophysics/IMAPP, Radboud University Nijmegen,
              P.O. box 9010, 6500 GL Nijmegen, the Netherlands\\
              \email{m.nielsen@astro.ru.nl}
         \and
             Sterrenkundig Instituut Anton Pannekoek, Universiteit van Amsterdam (UvA),
	     P.O. box 94249, 1090 GE Amsterdam, the Netherlands
	 \and
	     Institute for Astronomy, KU Leuven, Celestijnenlaan 200D, 3001 Leuven, 
Belgium 
             }

   \date{Received -; accepted -}

% \abstract{}{}{}{}{} 
% 5 {} token are mandatory
 
  \abstract
  % context heading (optional)
  % {} leave it empty if necessary  
  % context heading (optional)
  % {} leave it empty if necessary  
   {The progenitors of supernovae type Ia are usually assumed to be either a single white dwarf accreting from a non-degenerate companion (the single degenerate channel) or the result of two merging white dwarfs (the double degenerate channel). However, no consensus currently exists as to which progenitor scenario is the correct one, or whether the observed supernovae Ia rate is produced by a combination of both channels. Unlike a double degenerate progenitor a single degenerate progenitor is expected to emit supersoft X-rays for a prolonged period of time ($\sim$ 1 Myr) as a result of the burning of accreted matter on the surface of the white dwarf. An argument against the single degenerate channel as a significant producer of supernovae type Ia has been the lack of observed supersoft X-ray sources and the lower-than-expected integrated soft X-ray flux from elliptical galaxies.}
  % aims heading (mandatory)
   {We wish to determine if it is possible to obscure the supersoft X-ray emission from a nuclear burning white dwarf in an accreting single degenerate binary system. In case of obscured systems we wish to determine their general observational characteristics.}
  % methods heading (mandatory)
   {We examine the emergent X-ray emission from a canonical supersoft X-ray system surrounded by a spherically symmetric configuration of material, assuming a black body spectrum with $T_{bb}=50$ eV and $L=10^{38} \mathrm{erg} \cdot \mathrm{s}^{-1}$. The circumbinary material is assumed to be of solar chemical abundances, and we leave the mechanism behind the mass loss into the circumbinary region unspecified.}
  % results heading (mandatory)
   {We find that relatively small circumstellar mass loss rates, $\dot{M}=10^{-9}-10^{-8} \mathrm{M}_{\odot}\mathrm{yr}^{-1}$, at binary separations of $\sim 1$ AU or less, will cause significant attenuation of the X-rays from the supersoft X-ray source. Such circumstellar mass loss rates are sufficient to make a canonical supersoft X-ray source in typical external galaxies unobservable in Chandra.}
  % conclusions heading (optional), leave it empty if necessary
   {If steadily accreting, nuclear burning white dwarfs are canonical supersoft X-ray sources our analysis suggests that they can be obscured by relatively modest circumbinary mass loss rates. This may explain the discrepancy of supersoft sources compared to the supernova Ia rate inferred from observations if the single degenerate progenitor scenario contributes significantly to the supernova Ia rate. Recycled emissions from obscured systems may be visible in other wavebands than X-rays. It may also explain the lack of observed supersoft sources in symbiotic binary systems.}

   \keywords{(Stars:) supernovae: individual - (Stars:) binaries: close - Accretion, accretion disks - (Stars:) white dwarfs - Stars: winds, outflows - X-rays: binaries}

   \maketitle

%
%________________________________________________________________

\section{Introduction} \label{sect:Introduction}

Type Ia supernovae (SNe) are believed to be carbon-oxygen white dwarfs (WDs) close to the Chandrasekhar mass that undergo thermonuclear runaway in their centers. The resulting explosion produces radioactive iron-group elements, and the subsequent decay of these, most notably of $^ {56}$Ni, powers characteristic light curves that obey a well-known relation between luminosity at maximum light and fall-off time (Phillips \cite{Phillips.1993}). As a result, SNe Ia are considered standardizable cosmological candles. To this effect they have been utilized to suggest that the expansion of the Universe is accelerating (Riess et al. \cite{Riess.et.al.1998}, Perlmutter et al. \cite{Perlmutter.et.al.1999}), which in turn has given rise to the paradigm of Dark Energy. Additionally, the energy release of SNe Ia is large enough to influence the dynamics of their host galaxies, and the nucleosynthesis taking place during the explosions is the main source of iron group elements in galactic chemistries.

Despite decades of intense research on the subject, the exact nature of the progenitor systems of these important astrophysical explosions remains unclear. Carbon-oxygen WDs are characteristically formed at masses much lower ($\sim 0.6$ M$_{\odot}$) than that needed for thermonuclear runaway ($\sim 1.37$ M$_{\odot}$), and there is no known process by which an isolated sub-Chandrasekhar mass WD can grow to the critical mass at which it explodes as a SN Ia. Hence, it is usually agreed that SNe Ia can only arise in binary systems, where a WD accretes matter from a companion star. However, the exact method of accretion remains disputed. Two binary progenitor scenarios are usually considered: the single degenerate (SD), in which a WD accretes mass from a non-degenerate companion (main sequence or giant star, Whelan \& Iben \cite{Whelan.Iben.1973}), and the double degenerate (DD), in which two sub-Chandrasekhar mass WDs merge with a mass at or above the mass needed to explode as a SN Ia (Webbink \cite{Webbink.1984}, Iben \& Tutukov \cite{Iben.Tutukov.1984}).

While the DD scenario has garnered considerable attention recently, the SD scenario has been the most popular scenario for a long time, and much work has been done on the physics of this progenitor scenario (e.g. Hachisu, Nomoto \& Kato \cite{Hachisu.et.al.1996}). It was shown by Nomoto (\cite{Nomoto.1982}) that steady nuclear burning of hydrogen-rich material accreted from a companion onto a massive ($\sim 1$ M$_{\odot}$) WD can only take place in a fairly narrow interval of accretion rates close to $10^{-7}$ M$_{\odot}\mathrm{yr}^{-1}$. At smaller or larger accretion rates it is unclear if the WD will be able to grow sufficiently in mass for a SN Ia to occur, due to possible mass loss from nova eruptions or stellar winds. This puts rather tight constraints on the parameters of the progenitor systems.

In the 1990's luminous supersoft X-ray sources (SSSs) were recognized as an important new class of X-ray source (Tr{\"u}mper et al. \cite{Truemper.et.al.1991}, Greiner et al. \cite{Greiner.et.al.1991}), based on observations of the Large Magellanic Cloud made with the Einstein Observatory in the late 70's and early 80's (Helfand \& Grabelsky \cite{Helfand.Grabelsky.1981}). Since then, newer generations of X-ray instruments such as \textit{ROSAT}, \textit{BeppoSAX}, \textit{XMM-Newton} and \textit{Chandra X-ray Observatory} have found similar sources in other galaxies, including the Milky Way. As the name suggests, SSSs are characterized by having a much softer spectrum than those of more commonly known X-ray binaries involving a neutron star or a black hole. Subsequently, van den Heuvel et al. (\cite{van.den.Heuvel.et.al.1992}) showed that a massive WD accreting from a companion star at the steady-burning rate will emit X-rays with a spectrum consistent with that observed for a certain subset of SSSs as a result of thermonuclear burning of the accreted material. This made SSSs interesting as possible SD progenitor systems of SNe Ia.

If WDs undergoing steady nuclear burning on their surfaces look like SSSs, and if the SD progenitor scenario is the dominant contributor to the SN Ia rate, then we should expect - at least naively - to see a corresponding population of SSSs large enough to account for the observed SN Ia rate. However, as recently pointed out, the observed number of SSSs (Di Stefano \cite{Di.Stefano.2010}) and integrated soft X-ray flux (Gilfanov \& Bogd{\'a}n \cite{Gilfanov.Bogdan.2010}) observed from external galaxies appear to be at least one and more likely two orders of magnitude too low to account for the SN Ia rate. Furthermore, pre-explosion observations of the positions of nearby ($\lesssim$ 25 Mpc) SNe Ia using archival Chandra data have so far yielded no detections (Nielsen et al. \cite{Nielsen.et.al.2012}). This dearth of SSSs could very well mean that the missing SSSs are simply not there, and hence that the SD progenitor scenario is not the dominant contributors to the SN Ia rate. An alternative possibility is that the nuclear burning WDs appearing as SSS in SD progenitors do in fact exist and produce a significant fraction of the total SN Ia rate, but are somehow hidden from view of our current observational capabilities (in X-rays) during much of their supersoft phase.

In the following we wish to explore the latter option. We consider a simple model of a massive, accreting WD with a companion star that is losing matter into the circumbinary region in addition to the matter it transfers to the accretor. The goal has been to determine how much cirumbinary material is needed to render a nuclear burning WD in a nearby galaxy undetectable as a SSS for a given combination of binary parameters.

A note on notation: in this paper we will refer to the rate of material that is lost into the circumbinary region simply as 'the mass loss rate', $\dot{M}$. This should not be confused with the rate of mass that is transferred to the WD accretor, $\dot{M}_{\mathrm{acc}}$. In our notation, the total rate of mass lost from the donor is $\dot{M}_{\mathrm{tot}} = \dot{M} + \dot{M}_{\mathrm{acc}}$

In section \ref{sect:Model} we describe our model, including the structure of the gas bubble surrounding the SSS and the contributions to the obscuration from neutral gas, ionized gas, and dust. In section \ref{sect:Results} we present the results of our calculations, and section \ref{sect:Observ.implic} discusses the observational implications of the results. Section \ref{sect:Discussion} discusses the caveats of our model, and section \ref{sect:Conclusion} concludes.

%
%________________________________________________________________

\section{Model} \label{sect:Model}

We consider the emergent radiation from a massive ($\sim 1\mathrm{ M}_{\odot}$) WD accreting mass from a companion star in a close binary system. The donor may be a main sequence or evolved star. The system is losing mass into the circumbinary region, and this mass loss has created a spherical distribution of matter (gas and possibly dust) around the binary which may absorb and/or scatter the X-rays. The mechanism behind the mass loss from the donor into the circumbinary region is left unspecified in our study, but may be envisioned to be e.g. a stellar wind, wind Roche-lobe overflow (Mohamed \& Podsiadlowski \cite{Mohamed.Podsiadlowski.2007}), stellar pulsations of the donor, tidal interactions (e.g. Chen et al. \cite{Chen.et.al.2011}), a recently expelled common envelope, or any other process by which material can be deposited in the circumbinary region instead of being accreted by the WD. It is also possible that the material in the circumstellar region may originate from the WD if it emits a wind (Nomoto et al. \cite{Nomoto.et.al.1979}). Whether the above-mentioned mechanisms are actually capable of producing a spherical circumbinary configuration of matter is a question we do not enter into in this study.

The WD is burning accreted material at its surface at the steady burning rate mentioned earlier. The resulting luminosity is that of a typical SSS, i.e. $L_{\mathrm{bol}} = 10^{38}$ erg/s, and the spectrum is a black body with $kT_{bb}=50$ eV. Our focus is on X-ray observations, and as the companion is not expected to emit appreciably in X-rays, in observational terms our model system reduces to a single nuclear burning WD within a bubble of circumbinary material.

\begin{figure}[ht]
\centerline{\includegraphics[width=\linewidth]{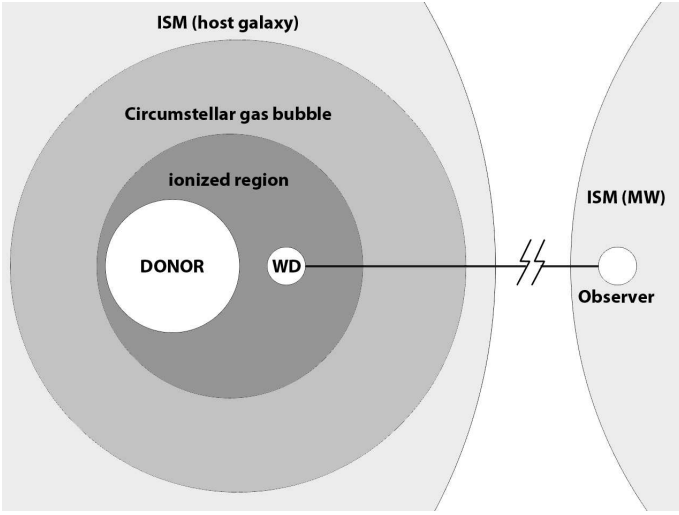}}
\caption{Schematic drawing of the model used in this study. The SSS system consists of a WD accreting material from a donor. The supersoft X-ray emission is the result of steady thermonuclear burning on the surface of the WD. The WD and donor are surrounded by a circumbinary configuration of gas and possibly dust. The radiation from the nuclear burning WD may ionize a region around the system; in this sketch the ionized region is localized narrowly around the binary, but for certain configurations the ionized region may extend to or beyond the edge of the circumbinary gas bubble. Before reaching an observer at Earth photons from the SSS pass through the circumbinary material, the ISM in the host galaxy, the IGM, and the ISM in the Milky Way. \ref{sect:Model}}
\label{fig:Model.schematic}
\end{figure}

\subsection{The gas bubble} \label{subsect:The.wind.bubble}
We parametrize the mass loss into the circumbinary region by a wind velocity, $u_w$, which we assume to be constant in the region outside of the position of the WD. We choose a value of $u_w=10$ km/s, typical for the winds of evolved intermediate mass stars (e.g. Panagia et al. \cite{Panagia.et.al.2006}).

As a first approximation, we assume the mass loss into the cirumbinary region to be spherically symmetric (see section \ref{sect:Discussion} for a discussion of the caveats of this assumption). The outer extent of the spherical distribution of material depends on the wind velocity and age of the mass losing phase of the donor star. For the chosen wind velocity the extent of the obscuring gas bubble is 2.1 AU$\mathrm{yr}^{-1}$ $\sim 10^{-5}$ pc$\mathrm{yr}^{-1}$.

\subsection{Obscuration by neutral gas} \label{subsect:Obsc.by.Neu.Gas}
In general, the optical depth $\tau$ along the line of sight between the source and the observer is the opacity $\kappa$ times the obscuring column $M$:
\begin{eqnarray}
 \mathrm{d}\tau=\kappa \rho dr \phantom{--} \Rightarrow \phantom{--} \tau = \kappa M , 
\end{eqnarray}
where $M = \int \rho dr$, and $\rho = \dot{M}/(4\pi r^2 u_{w})$, and $\dot{M}$ is the mass loss rate.

\begin{figure}[ht]
\centerline{\includegraphics[width=\linewidth]{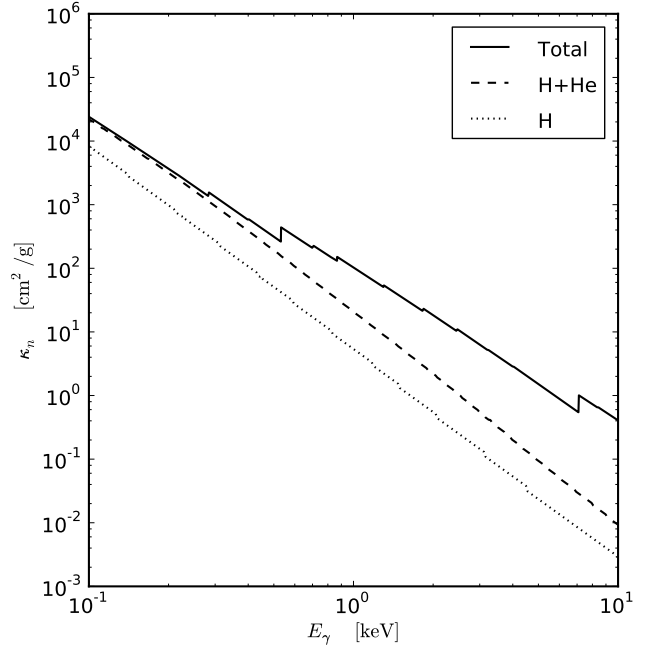}}
\caption{Opacities from K-shell ionizations for neutral gas and atomic metals as a function of photon energy. All values are for solar abundances. The dotted line gives the opacity of hydrogen alone, dashed is the opacity of hydrogen + helium, and solid is the total opacity of all chemical elements at solar abundances. Values are calculated from Morrison \& McCammon (1983). As explained in section \ref{sect:Discussion}, the relevant photon energies for Chandra are from 300 eV. Furthermore, for SSS we do not expect to see any photons above $\sim$ 3 keV.}
\label{fig:Cross.Sect}
\end{figure}

The total neutral column along the line of sight is the sum of the contribution from the local gas bubble, the ISM in the host galaxy, the IGM between the Milky Way and the host galaxy, and the ISM in the Milky Way. Therefore, for a given species of neutral gas in our spherically symmetric model the attenuation is formally given by
 \begin{eqnarray}
  \frac{I}{I_0} = \exp \Big( -\kappa_{\mathrm{n}} \Big( &&\frac{\dot{M}}{4\pi u_w} \Big( \frac{1}{r_0} - \frac{1}{r_1} \Big) \nonumber\\
					   &&+ \rho_{\mathrm{ISM}} (r_{\mathrm{host}}-r_1+r_{\mathrm{obs}}-r_{\mathrm{MW}}) \nonumber\\
					   &&+ \rho_{\mathrm{IGM}} (r_{\mathrm{MW}}-r_{\mathrm{host}}) \Big) \Big) \label{eq:Gas.Attenuation}
 \end{eqnarray}
where $\kappa_{\mathrm{n}}$ is the opacity of the neutral gas, $r_0$ is the inner radius of the neutral region of the gas species in question, $r_1$ is the outer radius of the spherical gas bubble, $r_{\mathrm{host}}$ is the distance from the source to the edge of the host galaxy, $r_{\mathrm{MW}}$ is the distance from the source to the edge of the Milky Way, $r_{\mathrm{obs}}$ is the total distance from the source to the observer, and $\rho_{\mathrm{ISM}}$ and $\rho_{\mathrm{IGM}}$ are the gas densities of the ISM and IGM, respectively. We put $r_0 = a$, where $a$ is the separation between the binary component. For both the wind material, the IGM, and the ISM we assume solar chemical abundances. X-ray absorption happens by way of K-shell ionizations, and the resulting photon energy dependent cross-sections are obtained from Morisson \& McCammon (\cite{Morrison.McCammon.1983}), as shown in figure \ref{fig:Cross.Sect}.

The formal expression in eq.(\ref{eq:Gas.Attenuation}) can be simplified somewhat. The contribution from the IGM is negligible, and we also assume that there is no significant contribution to the column from the host galaxy. This is routinely done for SSS studies (see e.g. Kahabka \& van den Heuvel \cite{Kahabka.van.den.Heuvel.2006}), and many of the galaxies used in the studies of for example Gilfanov \& Bogd{\'a}n (\cite{Gilfanov.Bogdan.2010}) and Di Stefano (\cite{Di.Stefano.2010}) are gas poor. Therefore, the column that we consider in our model is just the sum of the column of material in the Milky Way and the circumbinary material.

\subsection{Obscuration by ionized gas} \label{subsect:Obsc.by.Ion.Gas}
Close to the source the radiation generated by nuclear burning on the surface of the WD will photo-ionize the hydrogen in the gas. Since the peak energy of a canonical SSS is low ($kT_{bb} \sim 30-100$ eV) heavier elements are not likely to be appreciably ionized, and certainly not ionized to K-shell.

To determine the ionization structure of hydrogen around the source we follow the approach first suggested by Str{\"o}mgren (\cite{Stroemgren.1939}). The Str{\"o}mgren sphere is the volume around an ionizing source in which the ionization rate equals the recombination rate. The rate of recombinations to atomic energy level $n$ per unit volume is given by $ N_{R,n} = n_e n_p \beta_n(T_e)$, where $n_e$ and $n_p$ are number densities of free electrons and protons, respectively, and $\beta_n $ is the recombination efficiency of the $n$'th level, which depends on the electron temperature of the gas, $T_e$.

If we assume complete ionization within the ionized region then $n_e=n_p$ for hydrogen. Additionally, we can omit $\beta_1$ from the expression, since every recombination directly to the ground level emits a photon capable of causing another ionization which we assume it will do immediately. The total number of recombinations per unit time is then:

  \begin{eqnarray}
    N_{R\mathrm{,tot}} 	&=& \int_{r_0}^{R_I} n_e^2(r) \beta_{2+}(T_e) 4\pi r^2 dr \nonumber\\
			&=& 4\pi \beta_{2+}(T_e) \int_{r_0}^{R_I} n_e^2(r) r^2 dr \nonumber\\
			&=& \frac{\dot{M}^2 \beta_{2+}(T_e)}{4\pi u_w^2 m_g^2} \int_{r_0}^{R_I} r^{-2} dr \nonumber\\
  \end{eqnarray}
where $\beta_{2+} \approx 2\cdot10^{-10}T_e^{-3/4} \mathrm{ cm}^3\mathrm{/s}$ is the total recombination rate of transitions above the lowest ($\beta_{2+} = \sum_n\beta_n - \beta_1$), and we have used that
  \begin{eqnarray}
   n_e(r) = n_H(r) = \frac{\dot{M}}{4\pi r^2 u_w}\frac{1}{m_g}
  \end{eqnarray}
where $m_g$ is the mass of the hydrogen gas.

The number of ionizing photons emitted by the SSS per unit time is given by:
\begin{eqnarray}
 S_* = 4 \pi R^2 \int_{13.6\mathrm{ eV}}^{\infty} \frac{B_{\nu}}{h \nu}d\nu = \frac{L}{\sigma T^4} \int_{13.6\mathrm{ eV}}^{\infty} \frac{B_{\nu}}{h \nu}d\nu \label{S_*} \label{eq:number.ionizing.photons}
\end{eqnarray}
where $h$ is the Planck constant, $\sigma$ is the Stefan-Boltzman constant, $B_{\nu}$ is the frequency dependent Planck function, and $\nu$ is the frequency.

Setting $S_*$ equal to $N_{R\mathrm{,tot}}$ we get:
  \begin{eqnarray}
   \frac{\dot{M}^2 \beta_{2+}(T_e)}{4\pi u_w^2 m_g^2} \left( r_0^{-1} R_I^{-1} \right) &=& S_* \nonumber\\
   \Rightarrow r_0^{-1} - \frac{4\pi u_w^2 m_g^2 S_*}{\dot{M}^2 \beta_{2+}(T_e)} &=& R_I^{-1}   \label{eq:R_I_inverse}
  \end{eqnarray}

We assume that the temperature of the entire gas bubble is the effective temperature of the SSS. For $kT_{bb}=50\mathrm{ eV}$ the effective temperature is $5.8\cdot10^5\mathrm{ K}$. This is certainly an overestimation, but as discussed in section \ref{sect:Discussion} the impact of this inaccuracy is negligible.

Clearly, eq.(\ref{eq:R_I_inverse}) only has a physically meaningful solution if $r_0^{-1}$ is larger than the second term on the right-hand side. This gives us a constraint for the mass loss rate:
\begin{eqnarray}
 \dot{M} &>& \left( \frac{S_*4\pi u_w^2 m^2_g r_0}{\beta_2(T_e)} \right)^{1/2}   \nonumber\\
         &\equiv& \dot{M}_{\mathrm{Str}}. \label{eq:Crit.Mass.Loss.Rate}
\end{eqnarray}

For our model system, $L_{\mathrm{bol}}=10^{38}$ erg/s and $kT_{bb} = 50$ eV, the number of ionizing photons is $1.46\cdot10^{47}\mathrm{s}^{-1}$, and $\dot{M}_{\mathrm{Str}} = 2.00\cdot10^{-6}\mathrm{M}_{\odot}\mathrm{yr}^{-1}\cdot\left(\frac{r_0}{\mathrm{AU}}\right)^{1/2}$.

For mass loss rates larger than $\dot{M}_{\mathrm{Str}}$ there is a clearly defined inner ionized region, outside of which only neutral matter exists.

For mass loss rates smaller than $\dot{M}_{\mathrm{Str}}$ the expression for the ionized sphere will not be physically meaningful. In this case, the assumption that all photons capable of ionizing the gas are absorbed and cause ionizations is inaccurate. For such mass loss rates the gas bubble is not dense enough to absorb all photons capable of ionizing the gas, and there is no longer a clearly defined ionized region. Therefore, all hydrogen and helium in the gas bubble is fully ionized, and excess photons seep out into the interstellar medium, possibly causing further ionizations there.

In the ionized region, the only contribution to the obscuration from hydrogen and helium will be through Thomson scattering. The cross-section of Thomson scattering is largely indepent of photon energy, and therefore the free electrons produced in the ionizations will affect the absorption/scattering at all photon energies at approximately the same amount.

For X-ray binaries involving neutron stars or black holes the energies and densities involved may sometimes lead to Comptonization of the plasma, which has an impact on the obscuration caused by the Comptonized material. However, for the lower energies and densities involved in our model Compton scattering plays no role at all.

\subsection{Obscuration by dust}  \label{subsect:Obsc.by.Dust}
Depending on the properties and the mass loss mechanism of the donor star, a fraction of the metals in the gas may be condensed into dust grains, and we need to consider if the presence of dust changes the total X-ray absorption of the circumbinary material. 

As mentioned earlier, X-ray absorption happens by interaction of photons with K-shell electrons, and in this way the absorption cross section of an individual atom is largely independent of the location where the atom is found. However, putting atoms into grains represents a form of \textit{clumping}, which can decrease absorption if individual grains are already optically thick to X-rays of the considered energy. In this case, a part of the grain does not contribute to the X-ray opacity, because any X-ray photon will already be absorbed in the source-facing side of the grain. This can reduce the opacity by a so-called self-blanketing factor (Fireman \cite{Fireman.1974}) of

\begin{equation}
\label{eq:Self-blanketing}
f_{b} = \left( 1-e^{-\langle\tau_{\rm gr}\rangle} \right)/\tau_{\rm gr}
\end{equation}

Here $\langle \tau_{\rm gr} \rangle$ is the average optical depth of individual grains.  However, the effectiveness of this clumping of X-ray opacity is very limited. Even in cold diffuse clouds in the interstellar medium, important elements are hardly depleted into dust grains. For example, nitrogen and neon are abundant elements that remain entirely in the gas phase. In the mass loss flow of the donor star, this condensation will be even less complete.  If the donor star has an oxygen-rich chemistry, the entire carbon content of the wind and an equal amount of oxygen will be trapped in the very stable CO molecule and not participate in the condensation process (Gail and Sedlmayr \cite{Gail.Sedlmeyer.1986}). For solar-system-like element abundances (Anders and Ebihara \cite{Anders.Ebihara.1982}), this means that all of the carbon, all of the Nitrogen and Neon, and 40\% of the oxygen remain in the gas phase. As these are the most important absorbers for soft X-rays, it is clear that the soft X-ray absorption cross section will be reduced by a factor much less than 2.  Wilms et al. (\cite{Wilms.et.al.2000}) conclude that even in the ISM where condensation is rather complete, for grain sizes smaller than 0.3$\mu$m, the resulting effect will not exceed 10\%. In the donor mass flow, the effect will be even smaller. Therefore, we can safely ignore this complication and assume that all metals are in the gas phase, fully contributing to the X-ray absorption.

%
%________________________________________________________________

\section{Results} \label{sect:Results}

We first present the generic results of our model, i.e. results that do not depend on the instrument being used. Then we apply our results to a specific case, that of \textit{Chandra}'s ACIS-S detector.

\subsection{Generic results: Ionization structure and obscuration} \label{subsect:ionization.structure}
\begin{figure}[ht]
\centerline{\includegraphics[width=\linewidth]{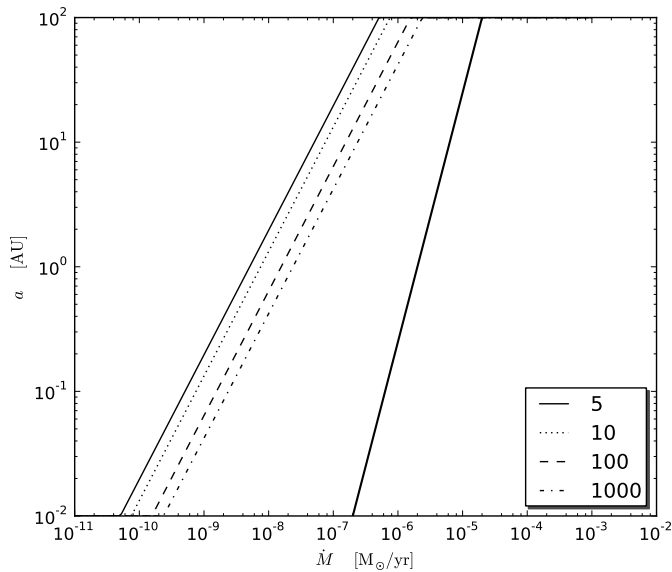}}
\caption{Contour lines showing the dependence of obscuration on mass loss rate and binary separation, for photons emitted at 350 eV. The left-most solid line corresponds to an attenuation factor of 5, the dotted line to an attenuation of 10, dashed to an attenuation of 100, and dot-dashed to an attenuation of 1000. The right-most solid line is the critical mass loss rate $\dot{M}_{\mathrm{Str}}$; to the left of this line the wind is not dense enough to sustain a clearly defined ionized region around the source. To the right of it a clearly defined ionized region exists around the source, while the material outside of this region is neutral.}
\label{fig:contourlines.loglog.350eV}
\end{figure}

Figure \ref{fig:contourlines.loglog.350eV} shows how the obscuration depends on orbital separation and mass loss rate for 350 eV photons emitted from a $L=10^{38}\mathrm{erg/s}$, $kT_{bb}=50 \mathrm{eV}$ SSS. Photons at 350 eV are safely above the photon energies at which typical X-ray observations are unreliable, but close enough to the peak of the black body curve that an unobscured SSS produces appreciable amounts of photons.

Our study shows that for binary separations around 1 AU a spherically symmetric mass loss rate of $\sim 10^{-8} \mathrm{M}_{\odot}\mathrm{yr}^{-1}$ is sufficient to fully obscure the supersoft X-ray emission from our model system.

It is evident from figure \ref{fig:contourlines.loglog.350eV}, for orbital separations around 1 AU a clearly defined ionized region will form at considerably larger mass loss rates ($\sim 10^{-6} \mathrm{M}_{\odot}\mathrm{yr}^{-1}$) than that needed for full obscuration. Hence,  at this photon energy, in between those mass loss rates the SSS will be fully obscured, but the binary will be surrounded by an extended ionized region that may be detectable at other wavelengths than X-rays (see section \ref{sect:Observ.implic}).

We note that the obscuration curves on fig.\ref{fig:contourlines.loglog.350eV} are for a photon energy of 350 eV. As the photon energy rises the curves move further to the right in the plot, while the curve for the critical mass loss rate remains in place (since the critical mass loss rate does not depend on the photon energy). This means that for larger photon energies one can imagine configurations for which these curves overlap, i.e. there is a clearly defined ionized region around the binary, while the X-ray emission from the binary is not fully obscured. However, this does not happen for our SSS at photon energies below $\sim 5$ keV, and the number of photons emitted by the system at these energies is miniscule.

\subsection{Specific results: Chandra's ACIS-S detector}
\begin{figure}[ht]
\centerline{\includegraphics[width=\linewidth]{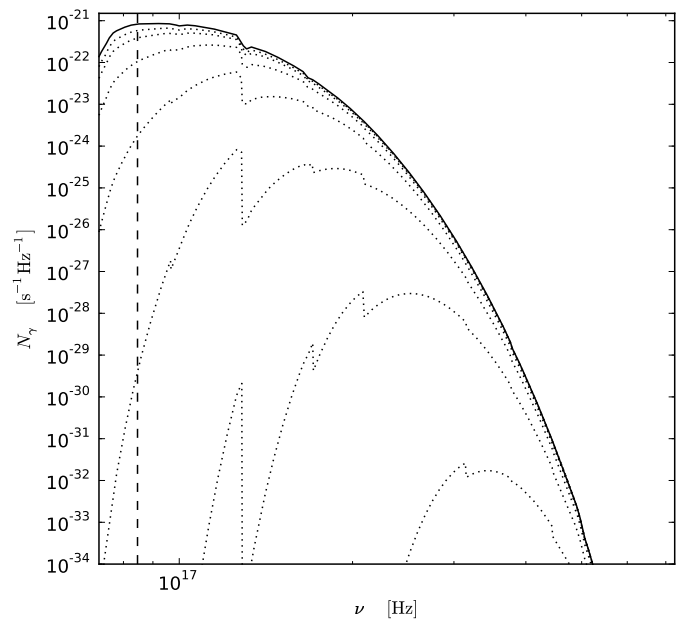}}
\caption{Unabsorbed and absorbed black body curves of a SSS in M101. The solid curve corresponds to a black body nuclear burning WD with $L=10^{38}$ erg/s, $kT_{bb}=50$ eV and no obscuring circumstellar matter (only galactic N$_{\mathrm{H}}$), folded with the effective area function of \textit{Chandra}'s ACIS-S detector. The dotted curves are the absorbed black body curves of the same source for seven logarithmically equidistant values of $\dot{M}$ from $10^{-9} \mathrm{M}_{\odot}\mathrm{yr}^{-1}$ to $10^{-6}$ M$_{\odot}\mathrm{yr}^{-1}$ (see table \ref{table:expected.no.of.photons}), with the mass loss rate increasing from left to right. The orbital separation and inner radius of the gas bubble is 1.5 AU. The numbers on each curve give the number of photons expected to be received in ACIS-S for an integration time of 40 ks. For comparison with figure \ref{fig:contourlines.loglog.350eV} the vertical dashed line is at a photon energy of 350 eV.}
\label{fig:absorbed.blackbody.plot}
\end{figure}

As an application, figure \ref{fig:absorbed.blackbody.plot} illustrates what our model system will look like if it resides in M101 and is observed with the ACIS-S detector on the \textit{Chandra} X-ray Observatory. The distance to M101 is taken to be 6.4 Mpc (Shappee \& Stanek \cite{Shappee.Stanek.2011}). The contribution to the column from the Milky Way is obtained from Dickey \& Lockman (\cite{Dickey.Lockman.1990}); for M101 we find a column of $1.15\cdot10^{20}$ N$_{\mathrm{H}}$/cm$^2$. We chose \textit{Chandra}'s ACIS detector as an example, since this is the instrument used by most groups (e.g. Voss \& Nelemans \cite{Voss.Nelemans.2008}, Roelofs et al. \cite{Roelofs.et.al.2008} Nelemans et al. \cite{Nelemans.et.al.2008}, Di Stefano \cite{Di.Stefano.2010}, Gilfanov \& Bogd{\'a}n \cite{Gilfanov.Bogdan.2010}, Nielsen et al. \cite{Nielsen.et.al.2012}).

\begin{table}
\caption{Attenuation ($N_{\gamma, \mathrm{obs}}/N_0$) of the integrated number of photons from our model system located in M101, as expected for different mass loss rate and observed with \textit{Chandra}'s ACIS-S detector. The mass loss rates listed correspond to the dotted lines in fig.\ref{fig:absorbed.blackbody.plot}. The integration has been done for photonenergies in the 0.3 to 3 keV range. For comparison, the number of photons expected in ACIS-S from the same source integrated over the same energy range in a completely empty (i.e. in the absence of galactic or circumbinary material) is $N_0 = 4.4\cdot10^{-5} \mathrm{s}^{-1}$. Note that the attenuation is for the integrated number of photons, and hence is not immediately comparable to fig.\ref{fig:contourlines.loglog.350eV}, as the values on that figure are for a single photon energy (350 eV).}
 \centering
  \begin{tabular}{c c c }
  \hline
  circumbinary $\dot{M}$		& $N_{\mathrm{obs}}/N_0$ \\
  $\mathrm{[M}_{\odot}\mathrm{/yr]}$	& \\
  \hline
  \hline
   $0$					& $9.0\cdot10^{-1}$ \\
   $10^{-9}$				& $8.0\cdot10^{-1}$ \\
   $3.1\cdot10^{-9}$			& $6.2\cdot10^{-1}$ \\
   $10^{-8}$				& $3.0\cdot10^{-1}$ \\
   $3.1\cdot10^{-8}$			& $4.6\cdot10^{-2}$ \\
   $10^{-7}$				& $6.7\cdot10^{-4}$ \\
   $3.1\cdot10^{-7}$			& $6.1\cdot10^{-7}$ \\
   $10^{-6}$				& $4.1\cdot10^{-11}$ \\
  \hline
\end{tabular} \label{table:expected.no.of.photons}
\end{table}

The photon energy dependent effective area function of ACIS-S can be found on the \textit{Chandra} homepage\footnote{http://cxc.cfa.harvard.edu/cgi-bin/build\_viewer.cgi?ea}. In order to find the number of photons detected we fold the calculated flux of the source with the effective area before integrating over all photon energies.

Since the launch of \textit{Chandra} in 1999 the sensitivity of the onboard detectors have degraded somewhat. We adopt the effective area function for the earliest possible \textit{Chandra} observations (cycle 3). If a source can be obscured sufficiently to be unobservable to the ACIS detectors when these were new and at their most sensitive, then such a source would also be unobservable to the older, less sensitive ACIS detectors. We note that the detectors on \textit{Chandra} are not sensitive to photons below roughly 100 eV. Additionally, \textit{Chandra} detections at photon energies between 100 and 300 eV are known to be unreliable, and analyses should therefore filter out photons at energies below this threshold. This means that observations of SSS spectra are in fact only observations of the high energy tails of their spectra, since the spectral peaks are located far below the detection threshold of \textit{Chandra}.

The number of photons we expect to receive in the relevant energy range (300 eV - 3 keV) for observations with ACIS-S is given in table \ref{table:expected.no.of.photons} for each of the curves in figure \ref{fig:absorbed.blackbody.plot}.

%
%________________________________________________________________

\section{Observational implications} \label{sect:Observ.implic}
In the preceding sections we have laid out the setup of the model and our results. To put our results into an observational context we will now review the current evidence for the existence of circumstellar matter around SNe Ia, which would be a signature of the SD progenitor scenario. We will also discuss the possible observational implications of our results for SN Ia progenitors.

\subsection{Evidence for CSM around SN Ia progenitors}
Several studies have found evidence for circumstellar matter in SNe Ia explosions:

High-velocity features in early-time optical observations of SN2003du were interpreted by Gerardy et al. (\cite{Gerardy.et.al.2004}) as evidence for the interaction of the outer-most layers of the SN ejecta with a dense circumstellar shell of solar metallicity material created by mass loss from the progenitor system prior to the explosion.

In a study of SN Ia remnants DEM L238 and DEM L249 Borkowski et al. (\cite{Borkowski.et.al.2006}) found bright central X-ray emissions surrounded by fainter shells and interpreted this as remnants of circumstellar media around the progenitors that had been shocked to emission by the SN ejecta.

In a more recent study, Chiotellis et al. (\cite{Chiotellis.et.al.2011}) compared 2D model simulations with observations of the historical SN Ia SN1604, also known as Kepler's SN. The SN remnant shows a peculiar nitrogen-rich shell-like structure in optical images. Simulations by Chiotellis and collaborators assumed a mass loss of $10^{-7}-10^{-6} \mathrm{M}_{\odot}\mathrm{yr}^{-1}$ and wind speeds of 5-20 km/s, typical of thermally pulsating asymptotic giant stars. The observed shell-like features are reproduced by their simulations, and interpreted as a shocked interaction layer between the progenitor's wind-blown circumstellar bubble and the SN ejecta. If correct, their results are consistent with a SD progenitor emitting a stellar wind prior to explosion.

Another recent study by Patat et al. (\cite{Patat.et.al.2007}) found blue-shifted absorption features of the Na I doublet (5896 \AA{} and 5890 \AA{}) in optical spectra of SN2006X, which was interpreted as evidence for gas outflows from the progenitor system. Their results were generalized by Sternberg et al. (\cite{Sternberg.et.al.2011}) who found similar features in an unbiased sample of 35 SNe Ia, indicating that such features, while not demonstrably present for other types of SNe, may be characteristic of SNe Ia.

\subsection{Upper limits}
Conversely, a number of studies have looked for, and failed to find, evidence for circumstellar matter in SN Ia explosions. These non-detections have led to upper limits being placed on the possible mass loss from SN Ia progenitor systems in several wave-bands.

Using optical spectra of SN2001el, Mattila et al. (\cite{Mattila.et.al.2004}), found upper limits of $\dot{M} \lesssim 9\cdot10^{-6}$ and $5\cdot 10^{-5}\mathrm{M}_{\odot}\mathrm{yr}^{-1}$ for the progenitor system of the SN, for wind speeds of 10 and 50 km/s, respectively.

Studies aimed at finding radio emission from the interaction between SN ejecta and circumstellar matter have been undertaken by several groups. No direct detection has been made at this point, but upper limits are reaching interesting values. Using the VLA, Panagia et al. \cite{Panagia.et.al.2006} found upper limits of $\sim 10^{-6}\mathrm{M}_{\odot}\mathrm{yr}^{-1}$ based on observations of 27 SNe Ia. Their analysis extrapolates from the assumption that the process behind radio emissions from SNe Ia are similar to those of SNe Ib/c.

Using more recent radio observations, Chomiuk et al. (\cite{Chomiuk.et.al.2011}) analyzed EVLA observations of early SN Ia spectra. From their non-detections they found typical upper limits $\dot{M}/u_w \lesssim 10^{-7} \mathrm{M}_{\odot}\mathrm{yr}^{-1}/(100 \mathrm{km/s)}$ for most sources (private communication). In another recent article, Chomiuk et al. (\cite{Chomiuk.et.al.2012}) reports upper limits of $\dot{M}/u_w = (6\cdot10^{-10} - 3\cdot10^{-9}) \mathrm{M}_{\odot}\mathrm{/yr}/(100\mathrm{km/s})$ for non-detections of radio emission from SN2011fe, the closest SN Ia in 25 years. For the lower wind speed used in our model the upper limits are correspondingly smaller; hence, the upper limit on the wind mass loss rate of the donor becomes $10^{-8}\mathrm{M}_{\odot}\mathrm{yr}^{-1}$ for typical SD SN Ia progenitors, and $(6\cdot10^{-11} - 3\cdot10^{-10}) \mathrm{M}_{\odot}\mathrm{/yr}$ for SN2011fe.

\subsection{Observational predictions}
We note that except for the case of SN2011fe the upper limits found in the studies mentioned above are all larger than what we require for full obscuration of systems with binary separations of $\lesssim 10 \mathrm{AU}$. The upper limits found by Chomiuk and collaborators come closest to constraining our results, and if their general limits are correct then our model cannot explain obscuration of systems with binary separations larger than 1 AU. If their limits for SN2011fe are correct then the non-detection of X-ray emissions from that particular SN cannot be explained by obscuration from circumbinary material in our model for a giant donor, since that would require binary separations of $\lesssim 10^{-2} \mathrm{AU}$, effectively placing the WD within the envelope of the giant. A MS donor cannot be ruled out by these radio upper limits, neither can a WD wind. Hopefully, future observations will provide either a detection of the shocked region or stronger general constraints with which to compare our model.

Another important point is that the density (and thereby mass loss rate) is not the only important parameter when determining upper limits. Our analysis shows that even relatively small circumbinary gas bubbles are able to fully obscure the system: As can be seen from the $r_0$-dependence in eq.(\ref{eq:Gas.Attenuation}), the obscuration is caused mainly by material in a very narrow region around the SSS. The outer extent of the circumstellar gas bubble is essentially irrelevant, except for very small bubbles where $r_0$ is comparable in size to $r_1$. This means that even quite compact systems, say up to a gas bubble radius of $\sim$ 10 AU, would be capable of obscuring a SSS, provided the mass loss rate is large enough. It follows that the mass loss does not have to have been 'on' for very long to obscure the system in X-rays. We emphasize that this does not solve the problems raised by Di Stefano (\cite{Di.Stefano.2010}) and Gilfanov \& Bogd{\'a}n (\cite{Gilfanov.Bogdan.2010}), as the source still need to be 'on' for a significant period ($\sim 10^6$ years) to be able to grow significantly in mass. But it may explain why many SSS appear to be highly variable.

In general, observations to detect or constrain the mass loss rate of the progenitor systems of SNe Ia need to be performed a very short time after the SN explosion. The bulk of the SN ejecta moves at $\sim 10,000$ km/s, corresponding to roughly 6 AU per day. Therefore, the interaction shock from a nuclear burning WD shrouded in an fully obscuring gas bubble with a radius in the range of a couple of tens of AU is unlikely to be detected by anything but the very earliest (1-2 days after explosion) observations. Since radio appears able to supply the best upper limits the ideal observing scheme would be to obtain EVLA observations of SNe Ia within a day or less of the explosion.

If the companion in our model system is evolved we can disregard orbital separations less than $\sim 0.5-1 \mathrm{ AU}$, since that will be inside the outer layers of the companion. At such short separations the WD is more likely to spiral into the companion or cause the envelope to be expelled, rather than go through a stable period as a SSS. However, several studies argue against the possibility of giant companions. For example, using pre-explosion archival \textit{Hubble Space Telescope} (HST) images Li et al. (\cite{Li.et.al.2011}) ruled out a luminous giant or supergiant as the companion to SN2011fe, although a sub-giant companion could not be excluded by the data.

If the companion is a main sequence (MS) star the system can exist stably at much smaller orbital separations, and the mass loss rates required for obscuration are correspondingly lower, as evident from figure \ref{fig:contourlines.loglog.350eV}. But the mass loss rate from such systems may also be much lower than for systems containing evolved stars.

There is also the possibility that the mass loss from the system is caused by a wind from the WD itself. This could happen if the mass loss from the donor to the accretor is slightly larger than the maximal steady burning accretion rate. The accreted mass could be supplied by either a MS or evolved donor. If the accretion rate is larger than the steady burning rate, the WD will 'puff up' from the accretion and emit a wind of its own, while possibly still burning some of the material on its surface, see Nomoto et al. (\cite{Nomoto.et.al.1979}). The 'orbital separation' in figure \ref{fig:contourlines.loglog.350eV} should then instead be perceived as the difference in radius between the nuclear burning layer and the wind emitting layer of the WD. For such small separations full obscuration can be achieved even with fairly small ($10^{-11}-10^{-10} \mathrm{M}_{\odot}\mathrm{yr}^{-1}$) mass loss rates. If we envision a WD accreting at slightly above the maximal steady burning rate and emitting a weak, spherical wind of this magnitude such a source would be completely obscured in our model. Potentially of more general interest, if it could be shown that the accretion process feeding a steady burning WD is not 100\% efficient, or if the X-ray emitting surface of the WD loses a very small fraction of the accreted material to the circumstellar region while it burns the result could be significant or complete obscuration. Therefore, these WD wind scenarios could potentially explain the absence of a large fraction of the SSS that become type Ia SNe in the SD scenario.

A possible observational characteristic of our model could be H-$\alpha$ emission caused by recombination in the ionized gas bubble. This emission may be visible in archival optical images of nearby SNe Ia, and it would be logical to suggest a systematic archival search for this kind of emission in pre-explosion images at the position of nearby SN Ia progenitors. However, such a search of the HST archive was performed by Voss et al. (2012, in prep.), who found no evidence for optical counterparts for nearby SNe Ia progenitors, so it is unclear whether such a search is actually feasible. Another option is to search for H-$\alpha$ emission in regions where such emissions are not to be expected, i.e. outside of young, star-forming regions. It might be possible to make an analysis similar to the SSS 'counting' done by Di Stefano (\cite{Di.Stefano.2010}), but in H-$\alpha$ instead of X-rays. We note, however, that even though dust is unimportant for the obscuration of X-rays from our model system it could possibly obscure the H-$\alpha$ emissions, since dust is a more efficient absorber at optical wavelengths. Depending on the tenuousness of the outer parts of the gas bubble there may also be forbidden lines, but this depends heavily on the density of the gas.

\subsection{Symbiotics}
Our results may also explain why few classic symbiotic binary systems are not observed as SSS. Despite the fact that the WDs in symbiotic systems are expected to be massive and accreting mass at rates comparable to the steady burning region only three symbiotic SSS are currently known (SMC3, Lin 358 and AG Dra). Somewhat analogous to the evidence for circumbinary matter around SN Ia progenitors observations of outbursts from recurrent novae in symbiotic systems also indicate the presence of significant amounts of circumbinary material, providing absorbing columns large enough to fully obscure the systems in quiescence (e.g. Shore et al. \cite{Shore.et.al.1996}). The physical parameters of typical symbiotic systems are $L_{bol} = 10^3-10^4 \mathrm{L}_{\odot} = 4 \cdot 10^{35} - 4 \cdot 10^{36}$ erg/s and $a=2-5$ AU, while the red giant wind emitted by the donor in such systems is roughly $10^{-7} \mathrm{M}_{\odot}\mathrm{/yr}$, see e.g. Mikolajewska (\cite{Mikolajewska.2012}). If we assume that the spectra of this type of source is a black body comparable to a canonical SSS then Figure \ref{fig:contourlines.loglog.350eV} shows that even if the luminoisities of these systems were as large as the ones expected for the nuclear burning SD SN Ia progenitors they would be completely obscured. The fact that they are observed to be one to two orders of magnitude less luminous only serves to make them even easier to obscure.

%
%________________________________________________________________

\section{Discussion} \label{sect:Discussion}
As mentioned earlier, we use a number of simplifying assumptions in our calculations. Here we discuss the caveats introduced by these assumptions.

\subsection{Density profile}
The assumption of constant wind speed from the surface of the companion star is probably not correct. In reality, the wind is accelerated by a variety of processes until it reaches its terminal velocity, and this is not expected to happen until well beyond the orbit of the binary. Therefore, our wind speed is probably too large. If the wind is accelerated through the system and doesn't reach the constant value used in our simulations until some time later, i.e. further away from the source, the density close to the source will be larger. So, this assumption also underestimates the amount of obscuration.

Also, we have assumed a spherically symmetric distribution of matter around the binary. This is in general not what is observed in symbiotic systems, where disk-like structures in the orbital plane are expected, possibly accompanied by bipolar outflows (e.g. Solf \& Ulrich \cite{Solf.Ulrich.1985}; Corradi \& Schwarz \cite{Corradi.Schwarz.1993}; Munari \& Patat \cite{Munari.Patat.1993}). Overall, it is difficult to say whether the assumption of sphericity is likely to over- or underestimate the amount of obscuration. In case of a non-spherical structure, the obscuration in specific cases depends sensitively on the inclination to the sight-line of the observer. This uncertainty could be dealt with if we had a firm understanding of the symmetries of the matter in the relevant binary systems. Given the absence of that understanding, combined with the earlier mentioned fact that only the material very close to the system has a significant effect on the obscuration, we believe that our spherically symmetric model is a reasonable first approximation. Hopefully, further studies will provide a better understanding of the density structures of the systems in question.

\subsection{Temperature}
In our calculations we have a assumed a constant temperature throughout the entire gas bubble, i.e. the temperature of the surface of the WD. In reality, the temperature of the gas bubble will fall off with the distance from the emitting SSS. However, since even at a temperature corresponding to the $T_{bb}$ of the SSS the elements of importance to absorption and scattering in the observationally relevant interval, i.e. elements heavier than helium but lighter than iron, will not be fully ionized. Our simplified temperature assumption only plays a role for hydrogen and helium, which are unimportant absorbers at the photon energies where \textit{Chandra} is sensitive. We therefore estimate this effect to be negligible for our purposes.

\subsection{Dust and stellar winds}
As explained in section \ref{subsect:Obsc.by.Dust} dust appears to be fairly unimportant in connection with the possible obscuration of a SSS.

However, if the X-ray source manages to ionize a large region in the circumbinary gas, in this region dust formation will not be possible. If radiation pressure on dust is an important factor for driving the wind in the first place, introducing the X-ray source may have significant effects on the mass loss rate from the companion. If dust has formed already, it might be destroyed again, for example by thermal evaporation due to X-ray heating of the grains (e.g. Fruchter et al. \cite{Fruchter.et.al.2001}), by charge explosions due to massive photo-electric ionizations (ibid), or by thermal sputtering in hot gas (e.g. Tielens et al. \cite{Tielens.et.al.1994}). In this study we have not considered such effects in detail. Instead, we have used a fixed mass loss rate as a parameter of the model and disregarded the consequences for X-ray absorption if part of the heavier atoms are present in the form of dust grains.

\subsection{Metallicity}
By using the model of Morrisson \& McCammon (\cite{Morrison.McCammon.1983}) we assumed solar metallicity of the obscuring material (both the wind material and ISM). As mentioned, the most important obscuring elements for the photon energies accesible with \textit{Chandra} are elements heavier than helium but lighter than iron. For metallicities different from the one used in this study the obscuration will scale accordingly.

\subsection{Spectrum} \label{subsect:Spectrum}
Following the original work of van den Heuvel et al. \cite{van.den.Heuvel.et.al.1992} we have assumed the SSS to be a simple black body. This is probably not an entirely accurate description of a nuclear burning WD (e.g. Ness et al. \cite{Ness.et.al.2003}, Rauch \cite{Rauch.2003}, Rauch \& Werner \cite{Rauch.Werner.2010}), and by assuming a black body spectrum we may well overestimate the temperature of the actual surface of the WD. However, deviations of the actual spectra from that of a black body is negligible in this context, since we are analyzing obscurations of several orders of magnitude. In addition to this, the temperature dependence in our calculations is quite weak (e.g. the number of ionizing photons in the calculation of the extent of the ionized region depends on temperature like $\sim T^{-3/4}$). Observations of SSS have often assumed a black body spectrum (e.g. Greiner \cite{Greiner.2000}), so comparisons of such observations with our model will in a sense be consistent. We note that the question of WD atmospheres is not particularly well understood at this point, so a very detailed analysis with additional assumptions would not necessarily improve the applicability of our results.

%
%______________________________________________________________

\section{Conclusions} \label{sect:Conclusion}

To date, it has mostly been assumed that nuclear burning WDs in SD progenitor systems would be more or less 'naked'. Consequently, the absence of a large enough number of these sources have been seen as a problem for the SD model, as there seem to be too few of these sources to account for the observed SN Ia rate.

We have examined a model system of a canonical SSS embedded in a spherical circumbinary gas bubble. The mechanism behind the formation of the gas bubble has been left unspecified, but could be the result of e.g. a stellar wind from an evolved companion, wind-RLOF, pulsations of the donor, or tidal effects between the binary components.

We have shown that for a certain critical mass loss rate (e.g. $\dot{M} \sim 10^{-6} \mathrm{ M}_{\odot}\mathrm{yr}^{-1}$ for $a\sim 1 \mathrm{ AU}$) a clearly defined, narrowly situated, ionized region will form around the SSS. For systems with mass loss rates below this critical value the SSS will be surrounded by extended ionized regions that may extend into the ISM. 

Our results suggest that for systems with $a \sim 1 \mathrm{AU}$ quite modest circumbinary mass loss rates ($\sim 10^{-9}-10^{-8} \mathrm{M}_{\odot}\mathrm{yr}^{-1}$) are sufficient to significantly obscure the nuclear burning WD SSS. This mass loss is in addition to the mass the donor loses to the accreting WD. For wider systems larger mass loss rates are needed. Even at orbital separations on the order of 100 AU the mass loss rates required for significant obscuration ($\sim 10^{-7} \mathrm{M}_{\odot}\mathrm{yr}^{-1}$) are not unrealistic for some late red giants or asymptotic giant branch stars. However, if they exist such wide systems are unlikely to be able to supply the mass loss rate required for steady burning.

The mass loss rate required for the ionized region to become clearly defined is several orders of magnitude larger than the mass loss rates needed for total obscuration at the relevant photon energies ($\lesssim 1 \mathrm{keV}$). This means that SSS systems with sufficient mass loss rates to fully obscure the X-ray emission will have an extended region of ionized material surrounding it. According to our model, for binary separations $\sim 1 \mathrm{AU}$ mass loss rates between $10^{-8}$ and $10^{-6} \mathrm{M}_{\odot}\mathrm{yr}^{-1}$ produce such systems. While not observable as SSS, these systems may instead be observable in IR, radio or H-$\alpha$ as recombination nebulae. To the extent that they are available multi-wavelength archival searches of pre-explosion images at the positions of nearby SNe Ia may find such emissions, even if no sources have been found in archival Chandra images. We also propose that H-$\alpha$ emissions outside of young regions may be evidence of obscured SSS.

If a steady burning WD emits a small amount of the accreted material, on the order of $\sim 10^{-11}-10^{-10} \mathrm{M}_{\odot}\mathrm{yr}^{-1}$, our calculations show that they may be completely undetectable in X-rays. We have no way to determine whether steady burning WDs accrete at 100\% efficiency, but if it does not then even such a small amount of material will have important consequences for the X-ray signature of such objects.

The full obscuration constraints for binary separation and mass loss rate presented above are probably too strict. Our model examined an observational best-case scenario, and when simplifying assumptions have been made they have consistently been made to favor a minimal amount of obscuration. For these reasons, even lower mass loss rates may be sufficient to fully obscure more realistic systems.

Our results may have implications for the SD scenario for type Ia SNe. The fact that it is comparatively easy to hide the X-ray emissions from nuclear burning WDs may help to explain some of the 'missing' systems mentioned in the Introduction. If a significant fraction of the progenitor systems can be shown to be severely or completely obscured by circumbinary material originating in the binaries themselves then it would explain the discrepancy between the SN Ia rate and the low number of observed SSS systems and integrated X-ray luminosity of ellipticals.

Our study may also explain why so few symbiotic systems are visible as SSS, since typical symbiotic systems are embedded in a dense wind, and are less luminous than the expected SD SSS systems.

In future work we plan to include our model in population synthesis simulations to determine if systems with the parameters required for obscuration are produced in large enough numbers to account for a significant fraction of the 'missing' SSSs.

%
%______________________________________________________________

\begin{acknowledgements}
      This research is supported by NWO Vidi grant 016.093.305. The authors would like to thank the anonymous referee for constructive criticism and suggestions which have significantly improved the quality of this manuscript. Additionally, MN is grateful to Jan Kuijpers, Rosanne Di Stefano, Joanna Mikolajewska, and Laura Chomiuk for helpful comments and suggestions.
\end{acknowledgements}

\end{document}